\documentclass[aps,prb, twocolumn, superscriptaddress,showpacs,showkeys,floatfix]{revtex4-1}
\usepackage{graphicx}
\usepackage{dcolumn}
\usepackage{bm}
\usepackage{hyperref}
\usepackage{natbib}
\usepackage{float}
\begin{document}
\preprint{APS}
\title{Broadband and wide-angle nonreciprocity with a non-Hermitian metamaterial}
\author{David Barton III}
\affiliation{Materials Science and Engineering, Stanford University, Stanford, California, 94305, USA}
% \email{dbarton@stanford.edu}
\author{Hadiseh Alaeian}
\affiliation{Department of Physics and Astronomy, Northwestern University, Evanston, Illinois, 60208, USA}

\author{Mark Lawrence}
\affiliation{Materials Science and Engineering, Stanford University, Stanford, California, 94305, USA}

\author{Jennifer Dionne}
\affiliation{Materials Science and Engineering, Stanford University, Stanford, California, 94305, USA}
\email{jdionne@stanford.edu}
\date{\today}

\begin{abstract}
We theoretically demonstrate a non-hermitian metamaterial exhibiting broadband and wide-angle nonreciprocity. The metamaterial consists of planar metal-dielectric layers with a Parity-Time (PT) symmetric distrubtion of loss and gain.  With increasing loss and gain, the bandstructure and bandgap are strongly modified; further, the PT potential leads to distinct internal field distributions when illuminated from different sides.  Including nonlinearities arising from natural loss and gain saturation leads to nonreciprocal transmission in the visible over a 50 nm wavelength and $\pm$60$^{\circ}$ angular range. \end{abstract}
\maketitle

\section{Introduction}
Reciprocity, the invariance of input and output upon exchange of emitter and detector\cite{jackson2007classical}, places strict limitations on the functionality and efficiency of optical devices, from solar cells\cite{GreenNano} to optical communication\cite{miller2010optical} to one-way cloaking devices\cite{pendry2006controlling}. The basic building block of modern computation, the diode, is perhaps the most familiar nonreciprocal element with negligible transmission of electrons traveling across a junction in one direction.  This effect is derived for electrons from Pauli exclusion, such that differential doping of charge carriers can induce a potential barrier to inhibit transport in one direction.  The bosonic nature of photons precludes Pauli exclusion; light is instead bound by Lorentz reciprocity, in which all linear and time-invariant optical systems have reciprocal transmission.  Overcoming reciprocity by breaking one of these conditions is thus paramount to the development of next generation optical devices.

Faraday rotation has traditionally been used to circumvent reciprocity.  By applying a magnetic field to a magneto-optical material, an asymmetric permittivity tensor is generated which breaks time reversal symmetry.  To bring nonreciprocity the micron scale, new routes have been theoretically proposed and experimentally demonstrated which rely on time-dependent modulation of the refractive index\cite{lira2012electrically,wang2013optical}, as well as the use of nonlinearities in a variety of schemes\cite{fan2012all,guo2016chip,gallo2001all,scalora1994photonic,tocci1995thin,chang2014parity}.  However, these implementations generally rely on waveguides or high-quality factor ($>10^3$) resonators; accordingly, the angular and frequency range over which nonreciprocity is achieved is limited to a few degree angular range and a 1 nm spectral range. In order to realize a broader range of applications with nonreciprocity, such as free-space nonreciprocal optics, new routes must be considered which operate over wide angular and frequency spectra.

Non-hermitian systems offer a route towards nonreciprocal transmission.  By judiciously structuring the real and imaginary parts of the refractive index $n$ such that $n(\vec{r}) = n^*(-\vec{r})$, a Parity-Time (PT) symmetric structure is generated\cite{el2007theory,bender1998real,bender1999pt}.  This specific type of non-hermitian system is naturally asymmetric, exhibiting exotic, but still reciprocal, properties such as unidirectional invisibility\cite{ramezani2010unidirectional,lin2011unidirectional} and bloch mode excitation\cite{wang2017unidirectional}, as well as one-way polarization conversion\cite{lawrence2014manifestation}.  Nonlinearities in these structures have been used to demonstrate nonreciprocal transmission in waveguides\cite{zhao2016metawaveguide}, resonators\cite{peng2014parity}, and electrical circuits\cite{bender2013observation}.  However, these current embodiments are still limited to on-chip two or four port structures.

Here, we theoretically demonstrate a new route to broadband and wide angle nonreciprocity by combining nonlinear non-hermiticity with metamaterials, i.e. effective media whose optical properties are defined by periodic structuring of deeply subwavelength components.  Taking inspiration from plasmonic metamaterials, which have been used to show exotic properties like negative refractive indices\cite{verhagen2010three}, Veselago lensing\cite{xu2013all}, and sub-diffraction limited imaging\cite{fang2005sub}, we design a metamaterial which has an optical band gap in the visible.  By adding loss and gain in a PT symmetric fashion, this structure shows broad tunability in the band structure and band gap with the degree of non-hermiticity\cite{AlaeianPRA}. Ultimately, the bandgap completely disappears with increasing loss and gain.  Within this spectral range, we demonstrate nonreciprocal transmission by considering saturation effects which would arise naturally in the loss and gain.  This system is nonresonant and does not rely on waveguide structures; as such, we demonstrate nonreciprocity over a wide  wavelength (50 nm) and angular ($\pm$60$^{\circ}$) range.  This effect arises from asymmetric and directional modulation of the loss-gain parameter, which is confirmed by comparing the nonlinear metamaterial to a linear one with uniform loss and gain. Extracting an intensity-dependent effective loss-gain parameter for the entire metamaterial, we find the induced modification is at least twice as strong in the backward direction compared to the forward direction.
\section{Optical Properties of the linear metamaterial}
A schematic of the structure is shown in Figure 1a.  Metallic layers clad dielectric channels in a 5 layer metal-dielectric-metal pattern, of equal thickness (30 nm), giving a unit cell length ($\Lambda$) of 150 nm.  The metal is modeled with a lossless Drude model ($\omega_p = 8.85\times 10^{15} s^{-1}$) representing silver to highlight the underlying physics of the system; these results can be extended to systems with ohmic losses without significantly affecting the observed trends (See Supplementary figure S2).  We choose a high index (n = 3.2, near the refractive index of TiO$_2$\cite{devore1951refractive}) dielectric, and additionally include gain (-i$\kappa$) or loss (+i$\kappa$) in a PT-symmetric fashion within the dielectric layers.  The amount of loss and gain is characterized by the non-Hermiticity parameter $\kappa$.  Light is incident at an angle $\theta$ relative to the layer normal, and is TM polarized, such that the magnetic field of the incident light is always transverse to the layers, pointing in the y-direction.  Here, we define waves which first interact with the gain layer of a unit cell as propagating in the forward direction, while waves which interact with the loss layer first are propagating in the backward direction.

We solve for the bandstructure\cite{Russel1995confined} of the infinitely large metamaterial when $\kappa$ = 0 and 0.5 to understand the impact PT symmetry has on the optical properties of this metamaterial.  While $\kappa = 0.5$ corresponds to gain higher than most materials, this represents the extrema of bandgap tunability in this material.  As seen in figures 1b and c, a band gap throughout the entire light cone (indicated by the dotted line) exists for wavelengths between 450 and 550 nm; as such, no free-space photons may be transmitted through the material.  However, as $\kappa$ is increased to 0.5, the band gap decreases and ultimately disappears such that light incident from any angle may couple into propagating modes in the material, consistent with prior investigations of PT metamaterials\cite{AlaeianPRA}.  For frequencies within this range, transmission can thus be modulated from zero to unity or greater for a given wavelength by tuning the loss and gain within the structure.

We next investigate the transmission of a 10 unit cell metamaterial, which has a total thickness of 1.5 $\mu$m, approximately three times the wavelengths considered. Figure 2a shows the transmission of $\theta = 45^{\circ}$ light from free space through this material as a function of wavelength for non-Hermiticity parameters of 0 and 0.37.  For the passive metamaterial $(\kappa = 0)$, the bandgap is readily apparent as a 100 nm wavelength region of low transmission.  Upon increasing $\kappa$ to 0.3734, the bandgap completely disappears, and transmission near unity is seen throughout the entire range (dashed black line).  Considering a wavelength of 500 nm, we can observe the impact $\kappa$ has on the transmission (figure 2b).  For a passive ($\kappa = 0$) material, the transmission is $\approx$$10^{-7}$ and increases super-exponentially until a value of $\kappa = 0.3734$, where transmission is unity.  Here, the reflection from the forward direction is zero, while reflection from the backward direction is not (10$^{-5}$), corresponding to an exceptional point in the scattering parameters.  Beyond this value transmission stays at or greater than 1, but is no longer a monotonically increasing function of $\kappa$. 

At the exceptional point, the field intensity distribution in the material shows significant differences when illuminated from the forward or backward directions, as shown in Figure 2c.  We plot the normalized magnetic field intensity as it is continuous across boundaries.  When illuminated from the gain side (forward direction), the intensity pattern for each unit cell is approximately the same, with maximum normalized magnetic field intensities of approximately 8 at the interface between metal and dielectric.  Conversely, when illuminated from the loss side (backward direction), the field intensities are larger in magnitude than their corresponding value from the gain side, with a maximum in the center.  Indeed, intensities near the center of the material are enhanced by over a factor of 80 compared to the incident field. This asymmetry is a direct consequence of the asymmetric loss-gain distribution inherent in PT-symmetric structures.  Here, the material is perfectly impedance matched in the forward direction with no reflection, while a cavity mode is excited from the backwards direction.  Note that, while the internal field distributions and reflected intensities are different, the transmission when illuminated from either direction is the same, as this material is still reciprocal.  In order to break reciprocity, nonlinearities or some form of time modulation is still required.

\section{Nonreciprocity via loss and gain saturation}
Considering the asymmetry inherent in PT systems and nonresonant field enhancements inherited from plasmonics, nonlinearities are an attractive route towards nonreciprocity.  Since the internal field intensity is enhanced from the backward direction, the nonlinear response of this material will be more pronounced from this direction.  Further, since the loss-gain parameter $\kappa$ modulates the bandstructure over a 100 nm range, modifications to this parameter will have dramatic impact on the optical properties.  Loss and gain saturation arise naturally as an intensity-dependent nonlinearity, and have been studied in wavelength-scale PT systems previously\cite{ramezani2010unidirectional,liu2014regularization,lin2011unidirectional}.  In the linear regime, when no saturation effects are present, the permittivity of the dielectric regions are described as
\begin{equation}
\epsilon_r = 1+\chi_{TiO_2} + \chi_{PT}
\end{equation}
where $\epsilon_r$ is the relative permittivity, $\chi_{TiO_2}$ is the susceptibility of the host dielectric (and hence $1+\chi_{TiO2} = n^2$), and $\chi_{PT}$ is the susceptibility of the added loss or gain.  As the intensity in the dielectric increases, saturation modifies the contribution of the PT susceptibility as\cite{andresen2010effect}:
\begin{equation}
\chi_{PT}(|E|^2) = \frac{\chi_{PT}^\infty}{1+\frac{|E|^2}{|E_{sat}|^2}}
\end{equation}
Here, $\chi_{PT}^\infty$ is the linear susceptibility, $|E|$ is the electric field within the medium, and $|E_{sat}|^2$ is the saturation electric field intensity, a materials parameter.  Note that the plasmonic fields inherent in this system mean that the nonlinear permittivity of the dielectric layers will be a strong function of position through the layer; as such, the nonlinear metamaterial is in general non-Hermitian rather than PT symmetric when the field intensities within the structure are not symmetric.  For the dielectric layers defined by a refractive index $n$ and loss or gain component $\kappa$, the full description of the permittivity with saturation reads as:

\begin{equation}
\epsilon_r(|E|^2) = n^2 + \frac{-\kappa^2 \pm 2in\kappa}{1+\frac{|E|^2}{|E_{sat}|^2}}
\end{equation}
where the $\pm$ indicates whether the material is the loss or gain layer, respectively.  While the gain or loss susceptibility, and hence $\kappa$ is a strong function of position and intensity, we will show that this non-Hermitian material can be homogenized to a linear PT-metamaterial whose non-Hermiticity parameter is uniformly tuned by the incident intensity.

To understand the nonlinear response of this metamaterial, we perform simulations using a Finite Element solver with a commercially available software (COMSOL Multiphysics).  For all nonlinear simulations, we consider an initial $\kappa = 0.3734$.  First, we consider illumination of 500 nm light incident at $\theta = 45^{\circ}$ as a function of incident intensity normalized to the saturation intensity. Figure 3a shows the transmission through the metamaterial when illuminated in the forward or backward direction.  In both cases, unity transmission is seen at low intensities, corresponding to the linear and reciprocal metamaterial.  As the incident intensity increases, transmission is reduced regardless of illumination direction.  However, backward illumination leads to lower transmission, consistent with the enhanced internal intensities seen in the linear material.  At 3.5\% $|E_{sat}|^2$, transmission is below 10\% in the backward direction, while approximately 50\% in the forward direction.  Plotting the ratio between the forward and backward directions, we can see the transmission or isolation ratio as a function of incident intensity in Figure 3b.  As expected, low incident intensity leads to a transmission ratio of 1, where the material acts reciprocally.  As the incident intensity increases, the transmission ratio monotonically increases to approximately 6.5 at 3.5\% $|E_{sat}|^2$.  The inset visually shows the transmission difference by plotting the outgoing plane wave when illuminated in the forward or backward direction for the intensity indicated by the green star.

Figure 3c shows transmission from the forward or reverse direction as a function of wavelength ($\theta=45^{\circ}$) for 3 different incident intensities.  We choose to study wavelengths in a 50 nm range centered on 500 nm, as the transmission is relatively flat and near unity within this range in the linear regime.  At a relatively low incident intensity (0.35\% $|E_{sat}|^2$), transmission is maintained at approximately unity when incident in the forward direction, while a parabolic depression of the transmission is seen in the backward direction, an indication of the opening bandgap.  As intensity increases, this region of reduced transmission increases, and we see transmission uniformly decreases in both the forward and backward cases, while the transmission ratio throughout the wavelength range monotonically increases.  At the maximum intensity studied (3.5\%$|E_{sat}|^2$), the transmission ratio varies between 5 and 12 within the studied wavelength range.  Thus, nonreciprocal transmission is observed over at least a 50 nm bandwidth.  Broadband nonreciprocity is additionally observed at other incident angles (see supplementary figure S3).  Note that, in principle, this bandwidth is limited only by the initial bandgap size and the incident intensity on the structure.  Both increasing the number of unit cells and increasing incident intensity can enhance the transmission ratio and bandwidth of operation.

We additionally consider nonreciprocal transmission when the metamaterial is illuminated from differing angles.  Figures 3e and 3f show the transmission for the same incident intensities as 3c and 3d, but consider illumination of 500 nm light incident at an angle ranging $\theta = 0^{\circ}$ to $60^{\circ}$.  In general, transmission increases with increasing angle, corresponding to a greater degree of coupling to plasmonic modes.  As intensity increases, transmission tends to decrease at lower incident angles, but transmission greater than 1 is seen at 60 degrees for $3.5\%|E_{sat}|^2$ incident intensity.  Here, we additionally observe a transmission ratio which spans 3 to 18 with incident angle, meaning that this metamaterial generates a nonreciprocal response throughout a large angular range.  Because the wavelengths studied here are above the surface plasmon frequency and the refractive index is negative, this may lead to the development of novel nonreciprocal optical elements such as a nonreciprocal Veselago lens,\cite{veselago1968electrodynamics} given suitable gain materials.

\section{Nonlinear Metamaterial Homogenization}

To better understand the underlying mechanism of nonreciprocity, we consider the transmission and reflection as a function of power.  Remarkably, while the nonlinear metamaterial has in general a spatially varying permittivity in the dielectric, both the reflection and transmission coefficients for a given direction can be well correlated with a uniform and linear value of the loss-gain parameter $\kappa_{eff}$ (see supplementary materials for a discussion of reflection and transmission for the nonlinear and linear metamaterials).  In doing so, this metamaterial can be homogenized to an intensity- and directionally-dependent PT-symmetric metamaterial (for the specific illumination condition).  A schematic of this homogenization scheme is given in Figure 4a.  Figure 4b gives the retrieved $\kappa_{eff}$ from the forward or backward direction for 500 nm light incident at $\theta = 45^{\circ}$.  $\kappa_{eff}$ is determined by separately considering the reflection and transmission values, and both results are plotted for each direction.  In the forward direction, almost identical values of $\kappa_{eff}$ are retrieved from the reflection and transmission parameters.  The agreement is not as exact in the backward direction (within 5\% relative error), likely due to the induced cavity mode and strongly varying field intensity.  This homogenization, possible because of the metamaterial structure, means that the bandstructure is directly modified by the incident intensity, and a spatially varying material can be treated as an effective homogenized nonlinear material.  
\section{Conclusions}
In conclusion, we have demonstrated a new route to optical nonreciprocity which employs saturation effects in parity-time symmetric plasmonic metamaterials.  Intensity-dependent tuning of the band structure enables time-reversal symmetry breaking in a system with deeply subwavelength building blocks.  Transmission differences around an order of magnitude are achieved while also maintaining high transmission in one direction.  This effect is observed over a 50 nm wavelength range within the band gap of the material.  This effect relies on the vast tunability of optical properties from a non-hermitian potential.  While relying on kerr-type nonlinearities has limiations as a full isolator\cite{shi2015limitations}, this metamaterial could be used in many exciting contexts, from computation to one-wave cloaks to non-reciprocal Veselago lenses.  These results point towards a new method to achieve nonreciprocity with an optical path length approaching sub-micron scales over a broad frequency and angular range.  
\section{Acknowledgements}
The authors thank all Dionne group members for insightful feedback on the work.  This work is supported by a Presidential Early Career Award administered through the Air Force Office of Scientific Research (Grant No. FA9550-15-1-0006), the National Science Foundation Emerging Frontiers in Research Initiative (Grant No. 1641109), a Chevron Stanford Graduate Fellowship, and Northrup Grumman; all are gratefully acknowledged.

\bibliographystyle{unsrt}
\bibliography{main}
\newpage
\begin{figure}[H]
\end{figure}
\newpage
\begin{figure}[H]
\centering
\includegraphics[width=\columnwidth]{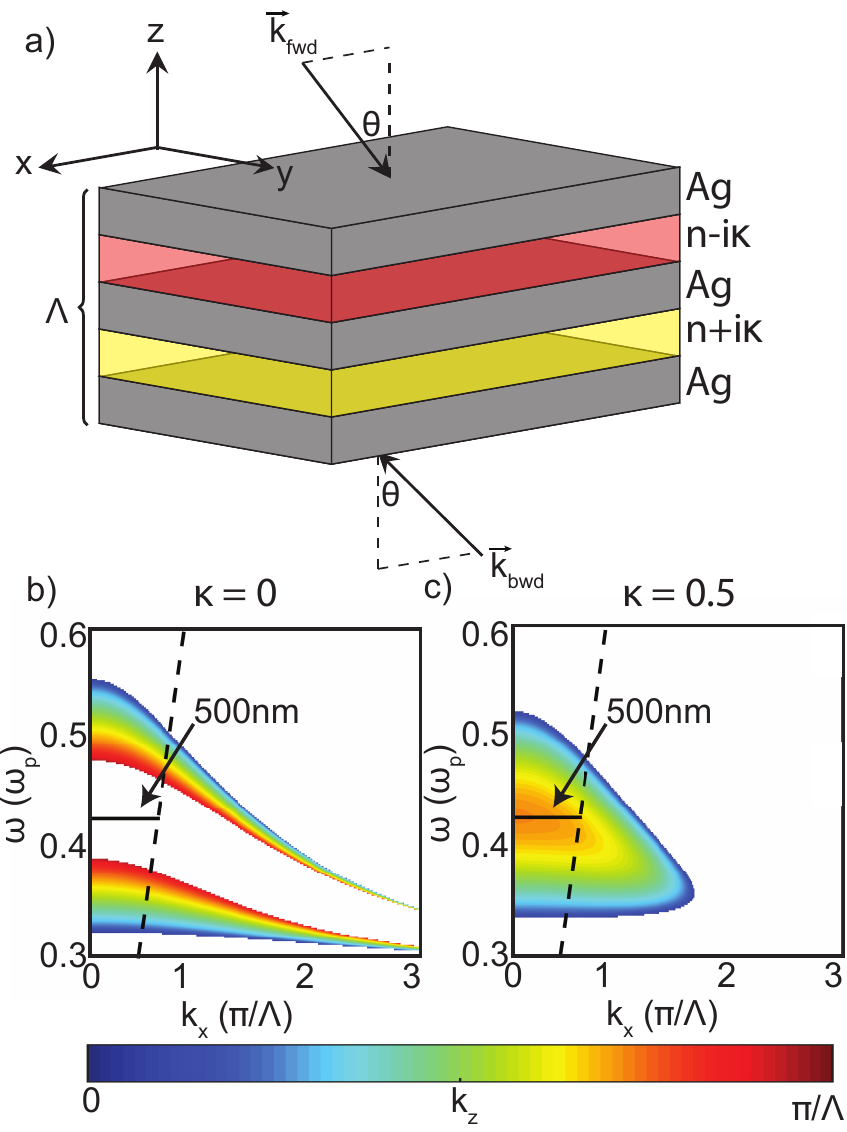}
\caption{Optical properties of non-hermitian metamaterial.  a) schematic of 5 layer unit cell.  Alternating layers of metal (Ag) and dielectric with either gain ($-i\kappa$) or loss ($+i\kappa$) b) bandstructure for TM polarized light ($H_y$) of metamaterial for $\kappa = 0$ and c) $\kappa = 0.5$, with 500 nm indicated with the horizontal line.  The freespace light line is indicated by the dashed black line.}
\end{figure}

\begin{figure}[H]
\centering
\includegraphics[width=\columnwidth]{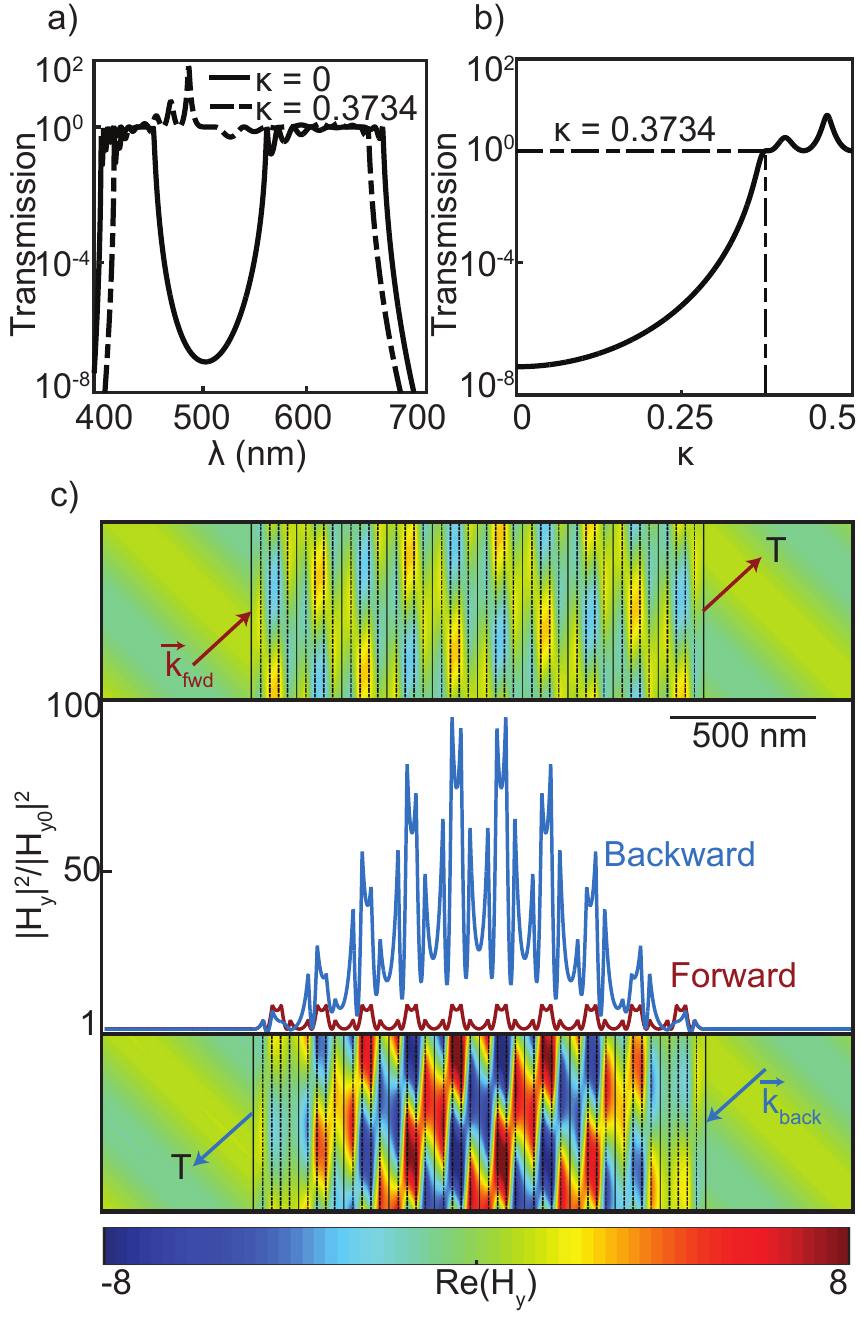}
\caption{Optical properties of 10 unit cell metamaterial.  a) Transmission as a function of wavelength for $\kappa = 0$ (solid) and $\kappa = 0.3734$ (dashed) with $\theta = 45^{\circ}$.  b) transmission as a function of  for 500 nm light incident at $\theta = 45^{\circ}$.  c) Magnetic field profile (normalized to incident field) for 500 nm light incident at $\theta = 45^{\circ}$  with $\kappa = 0.3734$ in the forward (top panel) and backward (bottom panel) illumination.  Middle: Directional magnetic field intensity enhancement.  Scale bar is 500 nm for all panels.}
\end{figure}

\begin{figure}[ht]
\centering
\includegraphics[width=\columnwidth]{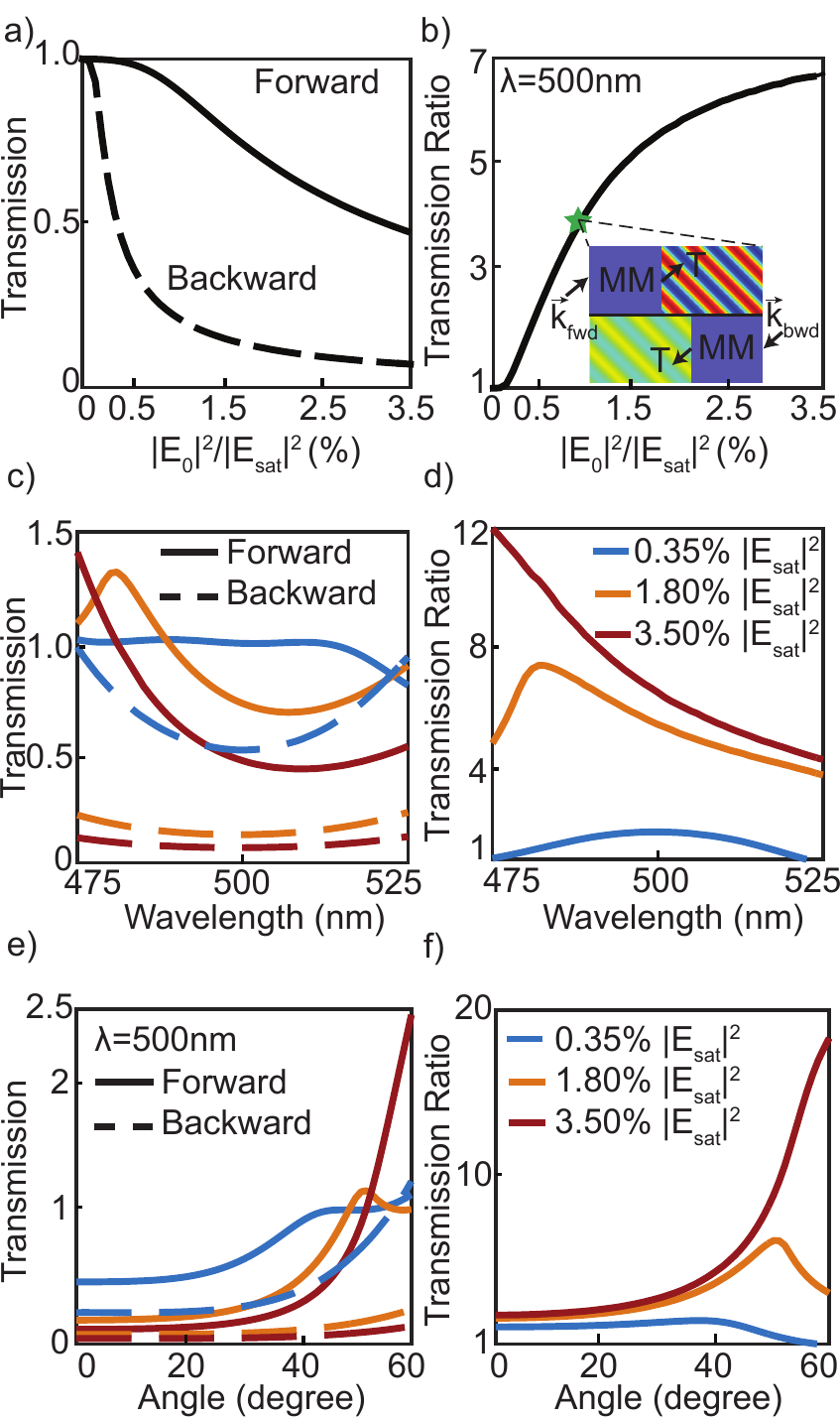}
\caption{Nonlinear and nonreciprocal transmission with initial $\kappa = 0.3734$.  a) Transmission and b) transmission ratio for 500 nm light as a function of normalized incident intensity.  c) Transmission and d) transmission ratio for 50 nm bandwidth around 500 nm for 0.35\% (blue), 1.8\% (orange), and 3.5\% (red) saturation intensity incident on material from forward (dashed) and backward (solid) directions.  Plane waves are incident for $\theta = 45^{\circ}$.  e) Transmission and f) transmission ratio of 500 nm light as a function of incident angle with the same intensities as in c) and d), from the forward (solid) or backward (dashed) directions.}
\end{figure}

\begin{figure}[H]
\centering
\includegraphics[width=\columnwidth]{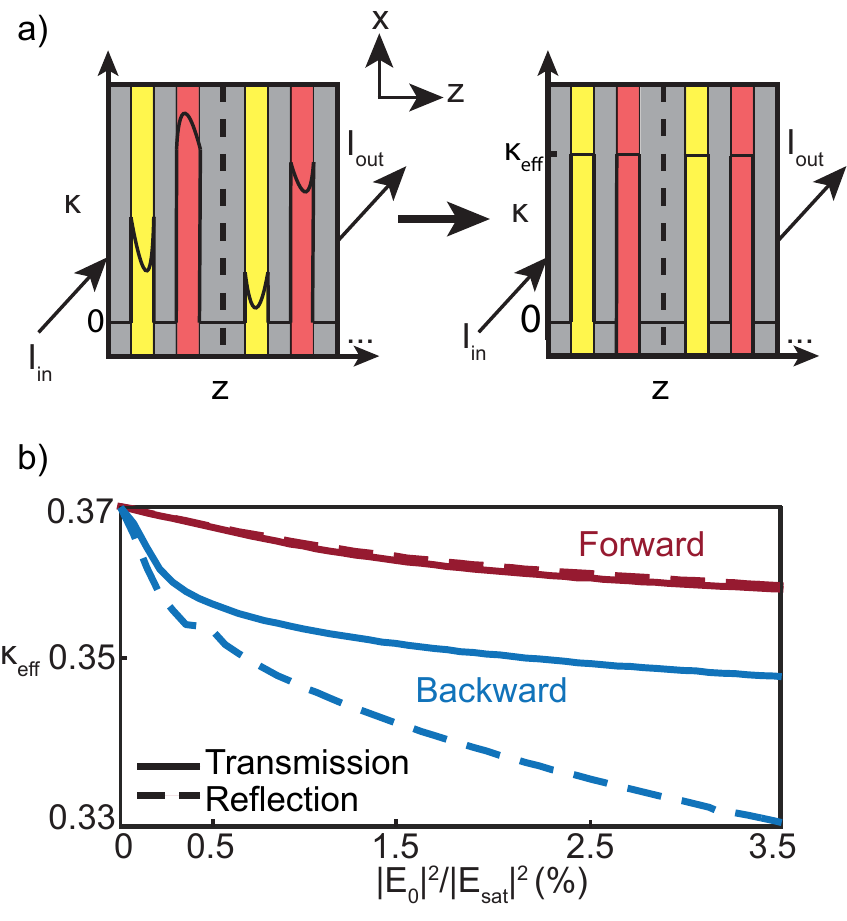}
\caption{Homogenization of nonlinear metamaterial.  a) schematic of homogenization.  b) retrieved $\kappa_{eff}$ for $\lambda = 500 nm$ and $\theta = 45^{\circ}$ as a function of incident electric field intensity in the forward (red) and backward (blue) directions.  Solid lines indicate retrieved value with transmission data, while dashed lines indicate retrieved parameter from reflection data.}
\end{figure}
\end{document}